\documentclass[12pt]{iopart}

\usepackage{graphicx}
\usepackage{amsfonts}
\usepackage{color}

\begin{document}

\title{Disassortative mixing accelerates consensus in the naming game}

\author{Han-Xin Yang$^{1}$ and Bing-Hong Wang$^{2}$}
\address{$^{1}$Department of Physics, Fuzhou University, Fuzhou
350108, P. R. China}
\address{$^{2}$Department of Modern Physics, University of Science and Technology of China, Hefei 230026, P. R. China}

\begin{abstract}

In this paper, we study the role of degree mixing in the naming
game. It is found that consensus can be accelerated on
disassortative networks. We provide a qualitative explanation of
this phenomenon based on clusters statistics. Compared with
assortative mixing, disassortative mixing can promote the merging of
different clusters, thus resulting in a shorter convergence time.
Other quantities, including the evolutions of the success rate, the
number of total words and the number of different words, are also
studied.
\end{abstract}



 \maketitle
\tableofcontents
\section{Introduction} \label{sec:intro}

Language dynamics, as an important issue in social
dynamics~\cite{social}, has been extensively studied with focusing
on the origins and evolution of
language~\cite{language1,language2,language3,language4,language5}.
To account for the emergence of shared vocabularies or conventions
in a community of interacting agents, a Naming Game (NG) model was
proposed~\cite{ng}. The NG has been widely applied in the study of
semiotic dynamics. A typical example is the so-called Talking Heads
experiment~\cite{exp}, in which a robot assigns names to objects
observed through cameras and negotiates with other robots about
these names. The NG can achieve the global consensus from a
multi-opinion state, which is apparently different from other
opinion models, such as the majority rule model~\cite{mr} and voter
model~\cite{vm}.

Recently, a minimal version of the NG based on principles of
statistical physics was proposed~\cite{mng}. This model simplifies
the original NG model but can as well reproduce the same
experimental phenomena. The minimal NG model has been studied in
fully connected graphs~\cite{full}, regular lattices~\cite{lattice},
complex networks (e.g., random networks, small-world networks and
scale-free
networks)~\cite{network1,network2,network3,network4,network5}, and
dynamic networks~\cite{dynamic}. Some modified versions of the
minimal naming-game model have been proposed to better characterize
the convergent behavior, such as connectivity-induced weighted
words~\cite{tang}, finite memory~\cite{wang}, local
broadcast~\cite{local1,local2,local3}, asymmetric
negotiation~\cite{yang}, reputation~\cite{reputation}, $n$
object~\cite{n}, and a preference for multi-word
agents~\cite{preference}.

Many real-world networks have various degree mixing
patterns~\cite{newman}: A network is said to show assortative mixing
if the nodes in the network that have many connections tend to be
connected to other nodes with many connections. A network is called
to show disassortative mixing if high-degree nodes tend to be
attached to low-degree ones. A measure of degree mixing for networks
is defined by the so-called assortativity coefficient~\cite{newman}:
\begin{equation}
r=\frac{M^{-1}\sum_{i}j_{i}k_{i}-[M^{-1}\sum_{i}\frac{1}{2}(j_{i}+k_{i})]^{2}}{M^{-1}\sum_{i}\frac{1}{2}(j_{i}^{2}+k_{i}^{2})-[M^{-1}\sum_{i}\frac{1}{2}(j_{i}+k_{i})]^{2}},
\end{equation}
where $j_{i}$, $k_{i}$ are the degrees of the nodes at the ends of
the $i$th edge, $M$ is the number of edges in the network, and $i =
1 . . .M$.  It was found that social networks are often
assortatively mixed ($r$ is positive), but that technological and
biological networks tend to be disassortative ($r$ is
negative)~\cite{newman}. For some celebrated network models, such as
the Erd\"{o}s-R\'{e}nyi random graphs and the Barab\'{a}si-Albert
(BA) scale-free networks~\cite{ba}, the assortativity coefficient
$r$ tends to be zero, indicating the lack of degree correlation. It
has been shown that the degree mixing plays an important role in
various dynamics such as the spread of epidemic~\cite{epidemic} and
the evolution of cooperation~\cite{cooperation}. However, the effect
of the degree mixing on the NG has not yet been studied in the NG.
In the following, we will show that disassortative mixing can
accelerate consensus in the NG.

\section{Model} \label{sec:model}

We firstly generate a scale-free network according to the
Barab\'{a}si-Albert (BA) model~\cite{ba}. Then we use the algorithm
proposed by Xulvi-Brunet and Sokolov (XS) to obtain networks with
expected degree mixing patterns~\cite{xs}: In order to get an
assortative network, each step randomly chooses two different edges
with four different ends, and then purposeful swaps the two edges by
linking the vertices with higher degrees and lower degrees,
respectively. By repeating this procedure forbidding multiple
connections and disconnected components, a network will become
degree assortativity without altering the degree distribution of the
original network. Through the opposite operation that one edge links
the highest and the lowest nodes and the other edge connects the two
remaining nodes, the network will become disassortative mixing.

After a network with the expected assortativity coefficient is
constructed, we play the NG. In the game, each node of a network
represents an agent. $N$ agents observe single object and try to
communicate its name with the others. Each agent is endowed with an
internal inventory to store a unlimited number of names. Initially,
each agent has an empty memory. Then the system evolves as follows:

(i) At each time step, a speaker $i$ is chosen at random and then
$i$ randomly chooses one of its neighbors $j$ as the hearer. This is
referred to be the directed NG~\cite{network2}.

(ii) If the speaker $i$'s inventory is empty, it invents a new word
and records it. Otherwise, if $i$ already knows one or more names of
the object, with equal probability it randomly choose one word from
its inventory. The invented or selected word is then transmitted to
the hearer.

(iii) If the hearer $j$ already has this transmitted word in its
inventory, negotiation is regarded as successful, and both agents
keep this common word and delete all other words in their
inventories; otherwise, the negotiation fails, and the new word is
included in the memory of the hearer without any deletion, i.e.,
learns the new word.

(iv) Repeating the above process until the consensus is reached,
that is, all agents have only one word in their inventories and
there are no different words in the system.

\section{Simulation results} \label{sec: results}

\begin{figure}
\begin{center}
\includegraphics[width=92mm]{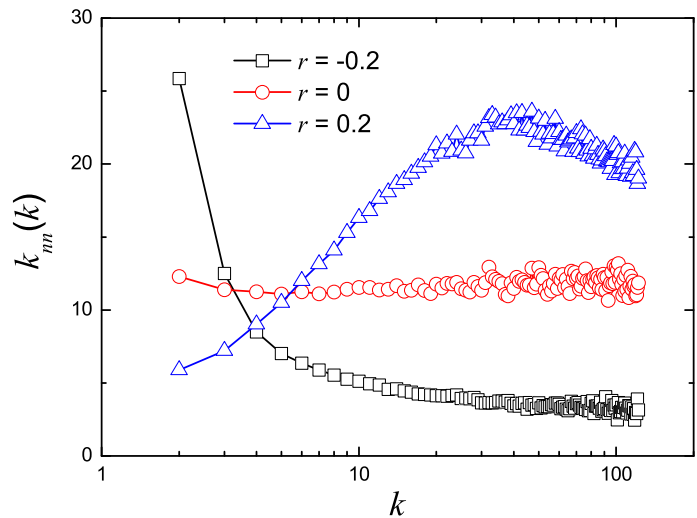}
\caption{(Color online) The average degree of nearest neighbors
$k_{nn}(k)$ as a function of degree $k$ for networks with different
assortativity coefficients $r$. The BA network size $N=5000$ and the
average degree $\langle k \rangle=4$. Each curve is an average of 50
different realizations.}\label{fig0}
\end{center}
\end{figure}

\begin{figure}
\begin{center}
\includegraphics[width=92mm]{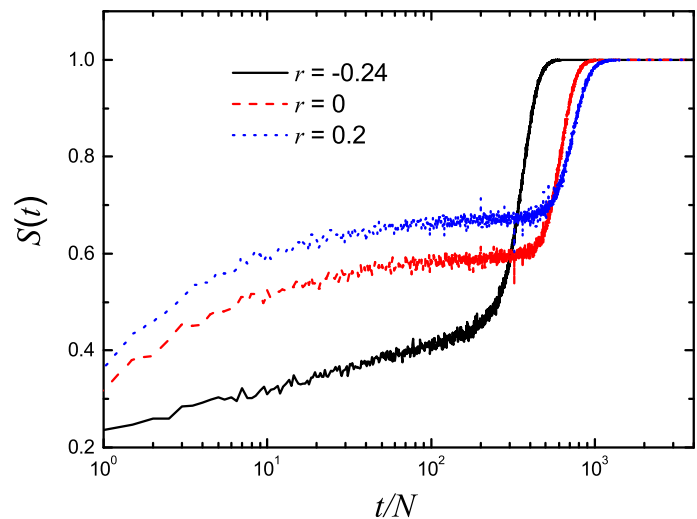}
\caption{(Color online) Evolution of success rate $S(t)$ for
different values of the assortativity coefficient $r$. The BA
network size $N=5000$ and the average degree $\langle k \rangle=4$.
Each curve is an average of 2000 different
realizations.}\label{fig1}
\end{center}
\end{figure}

Figure~\ref{fig0} shows the average degree of nearest neighbors
$k_{nn}(k)$ as a function of degree $k$ for networks with different
values of the assortativity coefficients $r$. From Fig.~\ref{fig0},
one can see that $k_{nn}(k)$ is almost independent of $k$ when
$r=0$. For assortative (disassortative) networks, $k_{nn}(k)$ is an
increasing (decreasing) function of $k$, indicating that the hubs
tend to connect with large(small)-degree nodes.

Figure~\ref{fig1} shows the evolution of success rate $S(t)$ for
different values of the assortativity coefficient $r$. From
Fig.~\ref{fig1}, one can see that at first $S(t)$ is lowest for
disassortative mixing ($r=-0.24$) and $S(t)$ is highest for
assortative mixing ($r=0.2$). But later on $S(t)$ increases most
quickly to 1 for disassortative networks, as compared to that for
assortative and uncorrelated ($r=0$) networks.

\begin{figure}
\begin{center}
\includegraphics[width=100mm]{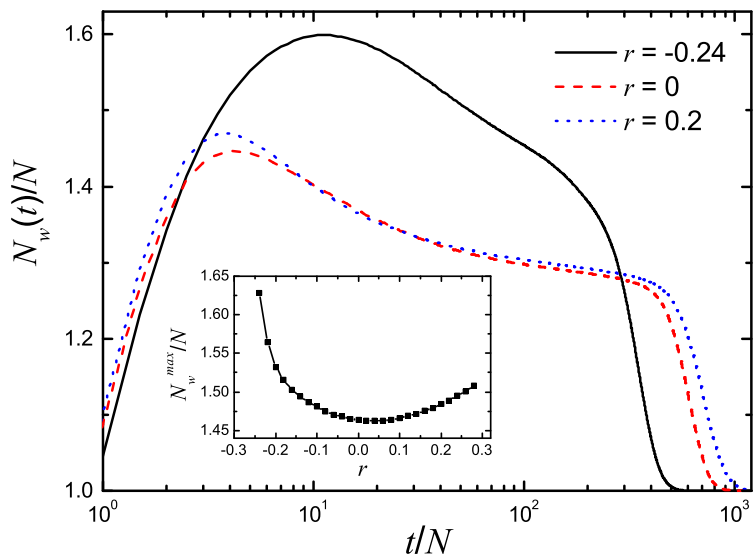}
\caption{(Color online) Evolution of $N_{w}(t)/N$, the average
number of words per agent, for different values of the assortativity
coefficient $r$. The inset shows the maximum total number of words
$N_{w}^{max}/N$ as a function of $r$. The BA network size $N=5000$
and the average degree $\langle k \rangle=4$. Each curve is an
average of 2000 different realizations. }\label{fig2}
\end{center}
\end{figure}

Figure~\ref{fig2} shows the average number of words per agent
$N_{w}(t)/N$ as time evolves. One can see that $N_{w}(t)/N$ grows
until it reaches a maximum (hereafter denoted by $N_{w}^{max}/N$),
and then it starts decreasing due to an increase in successful
interactions. The inset of Fig.~\ref{fig2} displays $N_{w}^{max}/N$
as function of the assortativity coefficient $r$. We find that
$N_{w}^{max}/N$ is not a monotonic function of $r$. In fact,
$N_{w}^{max}/N$ attains its minimum for $r\approx0.04$. The
non-monotonic relationship between the convergence time and the
maximum total memory was also found in Ref.~\cite{yang}.

\begin{figure}
\begin{center}
\includegraphics[width=92mm]{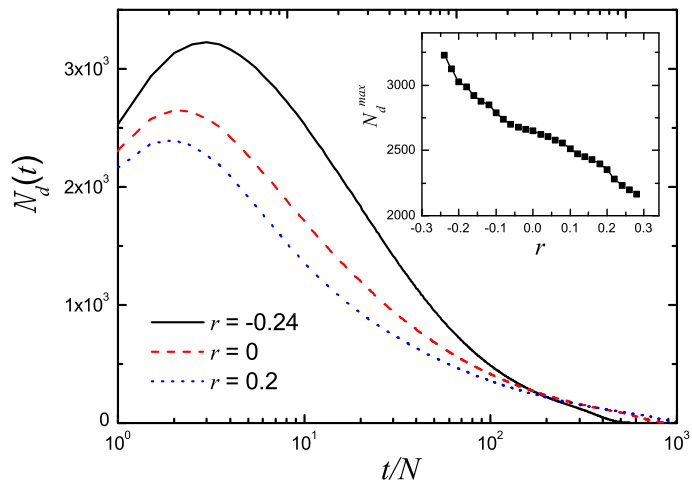}
\caption{(Color online) Evolution of the number of different words
$N_{d}(t)$ for different values of the assortativity coefficient
$r$. The inset shows the maximum number of different words
$N_{d}^{max}$ as a function of $r$. The BA network size $N=5000$ and
the average degree $\langle k \rangle=4$. Each curve is an average
of 2000 different realizations. }\label{fig3}
\end{center}
\end{figure}

Figure~\ref{fig3} shows the time evolutions of the number $N_{d}$ of
different words in the system. One can see that there exists a peak
of $N_{d}$ during the evolution for each value of assortativity
coefficient $r$. The inset of Fig.~\ref{fig3} shows that maximum
number of different words $N_{d}^{max}$ as a function of $r$. It is
found that $N_{d}^{max}$ decreases with the increase of $r$.

\begin{figure}
\begin{center}
\includegraphics[width=92mm]{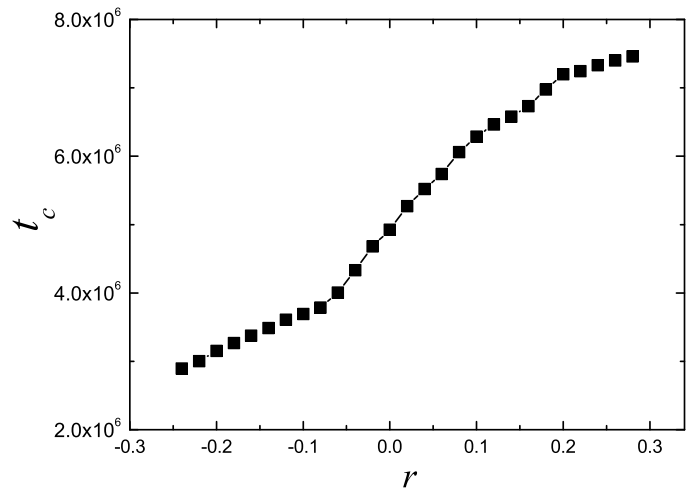}
\caption{Convergence time $t_{c}$ as a function of the assortativity
coefficient $r$. The BA network size $N=5000$ and the average degree
$\langle k \rangle=4$. Each data point is obtained by averaging over
2000 different realizations.
 }\label{fig4}
\end{center}
\end{figure}

Next we study the most important quantity, the convergence time
$t_{c}$ defined as the time steps for reaching the finial consensus.
Figure~\ref{fig4} shows $t_{c}$ as a function of the assortativity
coefficient $r$. One can see that $t_{c}$ increases with $r$,
indicating that disassortative mixing accelerates consensus in the
NG.

\begin{figure}
\begin{center}
\includegraphics[width=100mm]{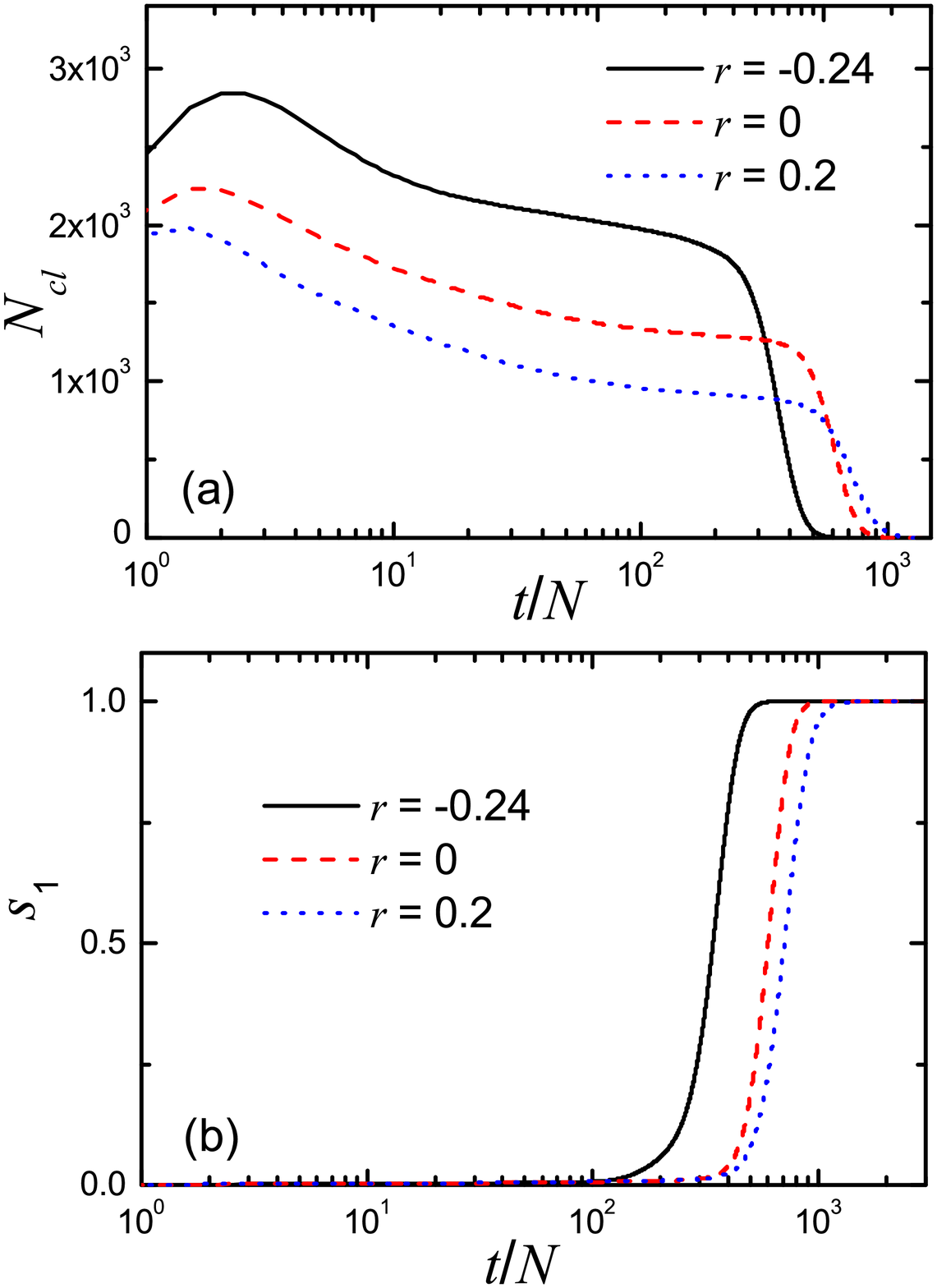}
\caption{(Color online) (a) The number of clusters $N_{cl}$ and (b)
the normalized size of the largest cluster $s_{1}$ as a function of
the rescaled time $t/N$ for different values of the assortativity
coefficient $r$. The BA network size $N=5000$ and the average degree
$\langle k \rangle=4$. Each curve is an average of 2000 different
realizations. }\label{fig5}
\end{center}
\end{figure}

\begin{figure}
\begin{center}
\includegraphics[width=95mm]{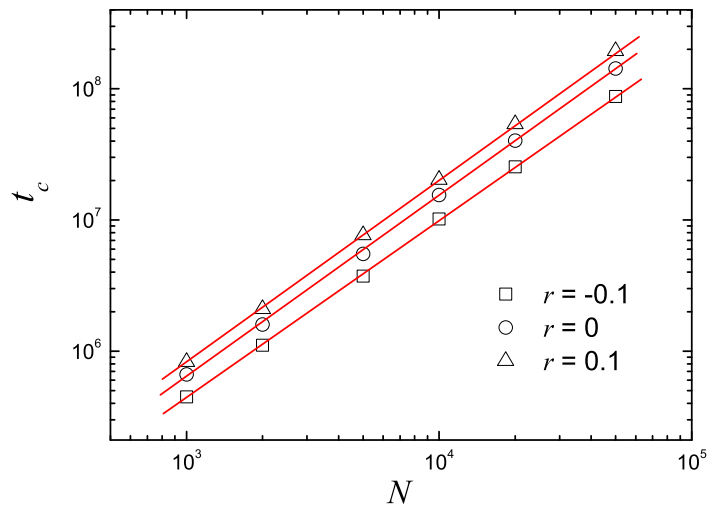}
\caption{(Color online) Convergence time $t_{c}$ as a function of
the BA network size $N$ for different values of the assortativity
coefficient $r$. The average degree $\langle k \rangle=4$. The slope
of the fitted line is 1.35, 1.38, 1.40 for $r=-0.1,0,0.1$
respectively. Each data is an average of 2000 different
realizations. }\label{fig6}
\end{center}
\end{figure}

\begin{figure}
\begin{center}
\includegraphics[width=95mm]{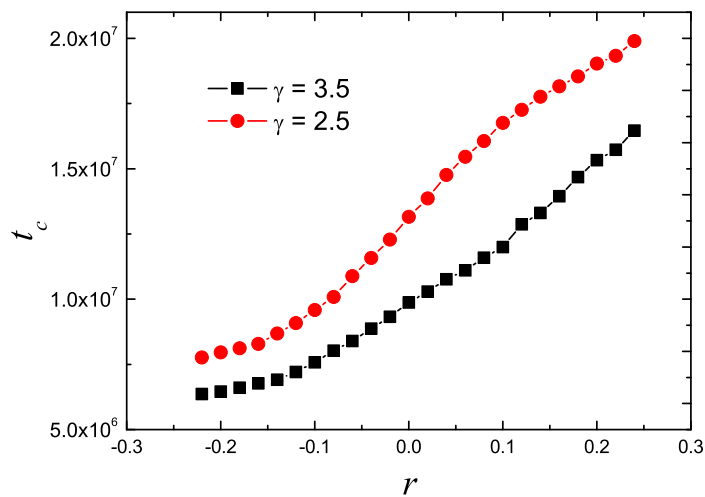}
\caption{(Color online) Convergence time $t_{c}$ as a function of
the assortativity coefficient $r$ for CM networks with different
values of the power-law exponent $\gamma$. The CM network size
$N=5000$ and the minimal degree $k_{min}=2$. Each data point is
obtained by averaging over 2000 different realizations.
}\label{fig7}
\end{center}
\end{figure}

\begin{figure}
\begin{center}
\includegraphics[width=95mm]{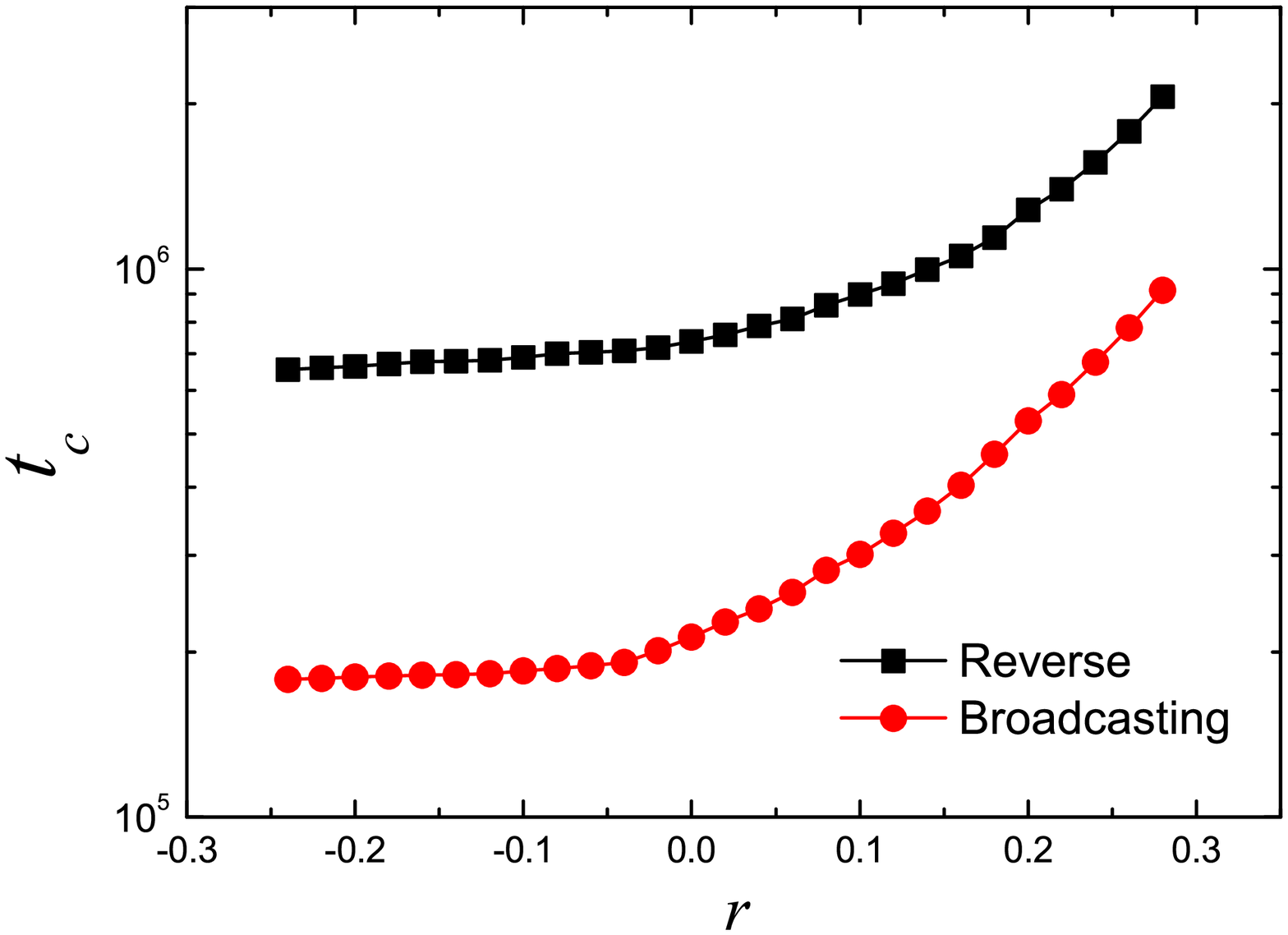}
\caption{(Color online) Convergence time $t_{c}$ as a function of
the assortativity coefficient $r$ in the reverse and broadcasting NG
respectively. The BA network size $N=5000$ and the average degree
$\langle k \rangle=4$. Each data point is obtained by averaging over
2000 different realizations. }\label{fig8}
\end{center}
\end{figure}

To understand the process of convergence to consensus, we study the
evolution of clusters of words. A cluster is a connected component
(subgraph) fully occupied by nodes sharing a common unique word. It
has been shown that the dynamics of the NG proceeds by formation of
such clusters~\cite{lattice}. Figure~\ref{fig5} shows the number of
clusters $N_{cl}$ and the normalized size of the largest cluster
$s_{1} =S_{L}/N$ as a function of the rescaled time $t/N$ for
different values of $r$, where $S_{L}$ is the size of the largest
cluster. From Fig.~\ref{fig5}(a), we see that the number of clusters
$N_{cl}$ reaches a plateau in the beginning, but then it rapidly
falls to one. Meanwhile, the normalized size of the largest cluster
$s_{1}$ remains very close to zero during the plateau, subsequently
it increases to one with a similar sudden transition [see
Fig.~\ref{fig5}(b)]. The sharp transitions for $N_{cl}$ and $s_{1}$
reflect the merging of different clusters. Previous studies have
shown that this merging accelerates consensus in the
NG~\cite{tang,yang} and other opinion models~\cite{yang1,yang2}.
From Fig.~\ref{fig5}, one can see that the merging of different
clusters is earlier and quicker in the case of disassortative mixing
($r=-0.24$), as compared to that in the cases of random mixing
($r=0$) and assortative mixing ($r=0.2$), thus leading to the faster
convergence.

Figure~\ref{fig6} shows the convergence time $t_{c}$ as a function
of the network size $N$ for different values of the assortativity
coefficient $r$. One can see that $t_{c}$ scales as $N^{\beta}$. The
value of exponent $\beta$ increases with $r$. For $r=-0.1$, 0, and
0.1, $\beta$ is about 1.35, 1.38 and 1.40 respectively. This result
manifests that disassortative mixing still accelerates consensus for
the large network size.

The above studies are conducted on BA scale-free networks which
follow a power-law degree distribution $P(k)\sim k^{-\gamma}$ with
$\gamma=3$. However, the finding that disassortative mixing
accelerates consensus is not restricted to BA networks. In fact, we
have observed a similar behavior for scale-free networks constructed
by the configuration model (CM)~\cite{cm}. Figure~\ref{fig7} shows
the convergence time $t_{c}$ as a function of the assortativity
coefficient $r$ for CM networks with different values of the
power-law exponent $\gamma$. From Fig.~\ref{fig7}, we also observe
that $t_{c}$ increases with $r$ for each value of $\gamma$.

Apart from the directed NG, there are other updating strategies of
NG, such as the reverse NG~\cite{network1} and the broadcasting
NG~\cite{local1}. In the reverse NG, we firstly choose the hearer at
random and then one of its neighbors as speaker. In the broadcasting
NG, a speaker transmits its word to all its neighbors at the same
time, rather than to a randomly selected one. It has been found that
the updating strategy greatly affects the process of convergence to
consensus~\cite{network2,local2}. From Fig.~\ref{fig8}, we also
observe that the convergence time $t_{c}$ increases with the
assortativity coefficient $r$ in the reverse NG and the broadcasting
NG.

\section{Conclusions and Discussions} \label{sec:discussion}

In conclusion, we have studied the impact of degree mixing on
consensus in the naming game. We adjust the assortativity
coefficient of a network by swapping the ends of the two edges while
keeping the degree of each node unchanged. We have found that the
convergence time decreases with the increase of the assortativity
coefficient. Compared with uncorrelated and assortative networks,
disassortative networks can accelerate the convergence to global
consensus. This finding is robust with respect to different types of
scale-free networks including the Barab\'{a}si-Albert model and the
configuration model, and to different kinds of updating strategies
including the directed, reverse and broadcasting naming games. We
have explained such phenomenon in terms of the evolution of the word
clusters. The merging of different clusters and the formation of big
clusters become easier in the case of disassortative mixing, as
compared to that in the cases of random mixing and assortative
mixing. We expect our work to provide new insights into agreement
dynamics on degree-correlated networks.

\section*{Acknowledgments}
This work was supported by the National Natural Science Foundation
of China (Grant Nos.£º11275186, 91024026, FOM2014OF001 and
71301028), the Natural Science Foundation of Fujian Province of
China (Grant No. 2013J05007), and the Research Foundation of Fuzhou
University (Grant No. 0110-600607).

\end{document}